\documentclass[twocolumn]{autart}    

\usepackage[dvips]{epsfig}    

\begin{document}

\begin{frontmatter}
\runtitle{Accretion of white dwarfs}  

\title{The accretion and spreading of matter on white dwarfs}

\thanks[footnoteinfo]{This paper was presented at Astronomy with Radioactives V. Corresponding author J.~L.~Fisker, Fax 1-574 631-5952}

\author[ND]{Jacob Lund Fisker}\ead{jfisker@nd.edu},    
\author[ND]{Dinshaw S. Balsara}\ead{dbalsara@nd.edu},               
\author[ND]{Tom Burger}\ead{tburger@nd.edu}  

\address[ND]{Department of Physics \& Joint Institute for Nuclear Astrophysics, University of Notre Dame, Notre Dame, IN 46556}


\begin{abstract}
For a slowly rotating non magnetized white dwarf the accretion disk extends all the way to the star. Here the matter impacts and spreads towards the poles as new matter continuously piles up behind it. We have solved the 3d compressible Navier-Stokes equations on an axisymmetric grid to determine the structure of this boundary layer for different viscosities corresponding to different accretion rates.
The high viscosity cases show a spreading BL which sets off a gravity wave in the surface matter. The accretion flow moves supersonically over the cusp making it susceptible to the rapid development of gravity wave and/or Kelvin-Helmholtz instabilities. 
This BL is optically thick and extends more than 30 degrees to either side of the disk plane after 3/4 of a Keplerian rotation period ($t_K$=19s).  
The low viscosity cases also show a spreading BL, but here the accretion flow does not set off gravity waves and it is optically thin.
\end{abstract}

\end{frontmatter}

\section{Introduction}\label{sec:introduction}
Cataclysmic variables (CV) is an interesting class of close binary stars comprising a hot white dwarf (WD) and a relatively lower mass red dwarf star filling its Roche lobe \cite{Warner95}. 
In such systems hydrogen-rich matter from the red dwarf exits through the inner Lagrange point and flows towards the white dwarf, where the matter proceeds to form an accretion disk around the white dwarf due to the excess angular momentum originating from the orbital motion of the binary \cite{Prendergast68}.

In the disk, turbulence \cite{Shakura73} and magnetic fields \cite{Balbus94a} dissipate potential energy and cause the matter to lose angular momentum. As a result the disk matter slowly falls inwards towards the white dwarf \cite{Pringle81}. 
If the WD is non-magnetic, the accretion disk extends all the way to the surface. 
Only half of the entire accretion luminosity, $L_{acc}=GM_*\dot{m}/R_*$, is emitted in the disk, since the matter is still moving at a roughly Keplerian velocity, $v_K\approx \sqrt{GM_*/R_*}$, before it is ultimately accreted \cite{Lynden-Bell74}.
The kinetic energy must therefore be converted into other forms of energy in order for the matter to come into co-rotation with the surface of the WD \cite{Pringle81}. 
This occurs in a comparably small (on the size of a disk scale height) boundary layer (BL) near the star.

Since the BL is much smaller than the disk itself and each radiates equal amounts of energy, the disk emits mainly in the optical and near UV, whereas the BL emits in the far UV and in soft X-rays. 
However, since the kinetic energy may also be converted into winds, WD rotation or heating, the details of the emitted spectrum depends on the detailed structure of the BL.
Therefore the structure of the BL is important to understand both to make the observational interpretations between the different components of the spectrum which is important in the study of dwarf novae and in the general interpretation of observational data, but also because it determines how the accreted matter ultimately distributes itself in the envelope of the WD, which is important for classical novae.
Before such a detailed analysis of the radiative transfer in conjunction with the hydrodynamics can be undertaken, it is important to understand the dynamical processes the lead to the formation of the BL. 
We focus on such a dynamical study in this paper leaving a radiation-hydrodynamical study for later.

Dwarf novae are a subclass of CVs in which a thermal instability in the accretion disk causes an increase in the matter transfer rate through the disk and thus an increase in the rate of gravitational energy release \cite{Cannizzo88}. This in turn causes a 2-5 mag increase in the disk luminosity and consequently in the BL luminosity. The BL may react in various ways due to enhanced accretion and detailed studies are required to determine its response \cite{Fisker05c}.

Another CV subclass is the classical nova. A classical nova obtains, when enough hydrogen-rich rich material has been accumulated in the envelope to reach the ignition point at the e-degenerate base of the envelope. The resulting thermonuclear runaway ejects part or all of the envelope. The initial CNO composition of the burning material should be strongly enhanced compared to the accreted material to account for composition of the observed ejecta \cite{Starrfield72}. Several mechanisms have been suggested for this CNO enhancement (see \cite{Jose05} and references therein). They can be roughly be divided into pre-burst mixing between accreted material and the underlying CO rich WD \cite{Kippenhahn78,MacDonald83,Rosner01,Alexakis04} by accretion driving instabilities or mixing with the underlying material when the convective zone of the thermonuclear runaway extends deep enough to dredge up CO material \cite{Starrfield72,Glasner97}. 

The BL has typically been solved in a model where the averaging of the vertical structure which is employed in accretion disk studies \cite{Shakura73} has been extrapolated to the surface. This reduces the description of the BL to a one dimensional problem in the radial direction \cite{Pringle81,Meyer82,Popham95,Collins98b}.
However, this assumption is not necessarily correct since the BL may spread out  \cite{Ferland82}.
These calculations were complimented with an analytic treatment of the meridional direction which showed that BL could spread and cover a significant part of the WD \cite{Piro04}.

Recognizing the multi-dimensionality of the problem several authors have attacked the problem directly using numerically methods culminating in simulations which included flux-limited radiative transport as well as different viscosities \cite{Robertson86,Kley87a,Kley89a,Kley91}. However, as these simulations predated those of \cite{Piro04} their setup did not include sufficient resolution in the radial and meridional planes to capture the dynamics and structure of the BL.

In this paper we set out to calculate the dynamical structure of the BL with sufficient numerical resolution to capture the dynamical evolution of the accretion flow and its interaction with the stellar surface (see \cite{Fisker05c}). 
Our model is presented in section~\ref{sec:model} and results for $\alpha=0.1$, $\alpha=0.03$, $\alpha=0.01$, $\alpha=0.005$, and $\alpha=0.001$ are given in section~\ref{sec:results}.

\section{Computational model}\label{sec:model}
The source of the angular momentum transport in the BL, which is responsible for transporting matter through the disk, is a combination of magnetic fields and turbulence. However, an a priori prescription, in particular in the BL, is a source of disagreement (see \cite{Popham95} and references therein). Instead the efficiency of the angular momentum transport can be parametrized with a coefficent, $\alpha$ \cite{Shakura73}.
Describing the angular momentum transport with a simple shear coeffienct means that the dynamics follows the Navier-Stokes equations. 
Here the Navier-Stokes equations as given by \cite{Mihalas84} are solved in spherical coordinates ($r$,$\theta$,$\phi$) on an axisymmetric mesh with 384 ratioed zones in the radial direction and 128 ratioed zones in the meridional range spanning 0 to 30 degrees from the disk plane -- same as \cite{Fisker05c}. Symmetry across the disk plane is assumed. This allows a doubling of the zone resolution at the same CPU-cost.

This allows us to resolve a pressure scale height in the radial direction and a disk scale height in the meridional direction for a stellar temperature of T=300000K and a disk temperature of T=1000000K. We assume a thin halo which consist of the same material as the disk (and the star). 
The halo provides a pressure that balances the pressure at the upper boundary of the disk, keeping the upper boundary of the disk in equilibrium.
The disk and halo model are described in \cite{Balsara04}.

The spatially and temporally second order agorithms in \verb+RIEMANN+ have been described in \cite{Roe96,Balsara98a,Balsara98b,Balsara99a,Balsara99b,Balsara04} and use many ideas from higher order WENO schemes (see \cite{Jiang96,Balsara00}) to reduce dissipation.
The matter in the model is subject to the central gravitational field of the underlying WD, which has a radius and mass of $R_*=9\times 10^8\,\textrm{cm}$ and $M_*=0.6M_\odot$ respectively -- same as \cite{Piro04}.
The model uses an ideal gas ($\gamma=5/3$) of a fully ionized solar composition ($\mu=0.62\,\textrm{g/mole}$) and assumes no radiation transport.

\section{Results \& Discussion}\label{sec:results}
In the next few subsections, we focus on several important aspects of the boundary dynamics. In section~\ref{subsec:ang}, we focus on the evolution of angular momentum in the boundary layer. In section~\ref{subsec:grav}, we discuss accretion based instabilities which may obtain and in section~\ref{subsec:optical} we relate the computations to observations.

\subsection{Angular momentum evolution}\label{subsec:ang}
As the simulations start, the shear force transfers angular momentum between differentially rotating disk annuli so that matter can move inwards and accrete on the star at the footpoint of the disk. 
Angular momentum transfer between the disk material and the stellar surface is also necessary so matter can move to higher latitudes. Otherwise the centrifugal barrier prevents matter from leaving the footpoint of the disk. It is this transfer which spins up the existing surface layers of the star. 
Angular momentum is also directly advected onto the star, because the orbiting disk material eventually accretes and forms the new surface.

\begin{figure}[thp]
 \epsfig{file=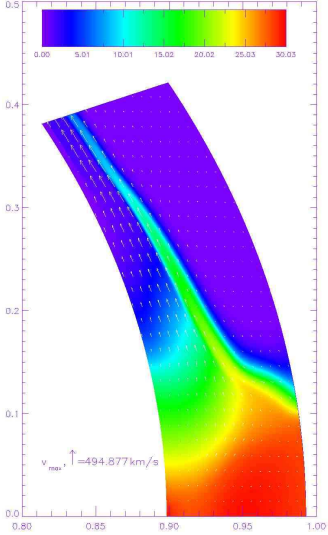,width=0.6\linewidth}
 \caption{A contour plot of the specific angular momentum, $\Omega=v_\phi/r$, for the $\alpha=0.1$ case after 3/4 of a Keplerian rotation period.}\label{fig:omega91}
\end{figure}

\begin{figure}[thp]
 \epsfig{file=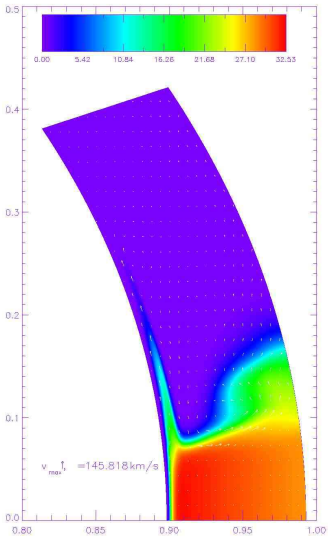,width=0.6\linewidth}
 \caption{A contour plot of the specific angular momentum for the $\alpha=0.001$ case after 3/4 of a Keplerian rotation period.}\label{fig:omega95}
\end{figure}

Fig.~\ref{fig:omega91} shows the resulting specific angular momentum after 3/4 of a Keplerian evolution for the $\alpha=0.1$ case, where fig.~\ref{fig:omega95} shows it for the $\alpha=0.001$ case. The halo is spun up as is the underlying star at the footpoint, where the disk connects. At higher lattitudes the spreading disk material is less dense and thus takes longer amounts of time to drag the star's surface around with it.


\begin{figure}[thp]
 \epsfig{file=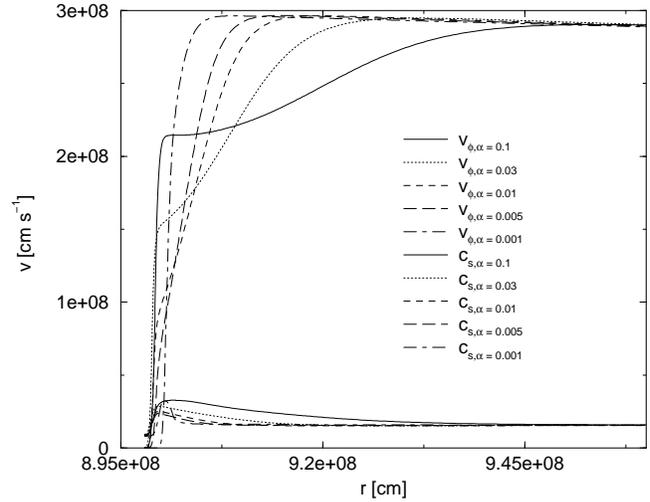, width=\linewidth}
 \caption{Toroidal ($\phi$) velocity at the meridional centerline of the disk shows the extent of the boundary layer as well as the sound speed after 3/4 of a Keplerian rotation period. Note that $v_\phi$ asymptotes to a higher value than $c_{s,\alpha}$.}\label{fig:vlza}
\end{figure}

Since degenerate matter is too ``slippery'',  turbulence and/or threaded magnetic fields might be the two most prominent ways to transfer a significant amount of angular momentum to the deeper layers of the WD \cite{Durisen73a}, where the relative motion drives the mixing between the accreted surface material and the deeper layers.
A supersonic component in the toroidal direction over most of the BL obtains for all values of $\alpha$ (see fig.~\ref{fig:vlza}).
Yet only the $\alpha=0.1$ case has a supersonic component in the poloidal directions, whereas the poloidal flows for $\alpha=0.03$, $\alpha=0.01$, $\alpha=0.005$, and $\alpha=0.001$ remain subsonic (see figs.~\ref{fig:omega91} and \ref{fig:omega95}) \cite{Fisker05c}. The supersonic velocities mean that the flows are susceptible to the rapid development of gravity wave and/or Kelvin Helmholtz instabilities. Such instabilities would develop self-consistently, if the interface had been given substantially more zones in which to form these instabilities. Due to our limited albeit high zone resolution, we can only provide the velocities which provide the driving mechanism for such instabilities. The calculation of the turbulent mixing which obtains from these instabilities must therefore be calculated using other models \cite{Kippenhahn78,MacDonald83,Rosner01,Alexakis04}.

Fig.~\ref{fig:vlza} shows the torodial velocity at the meridional centerline as a function of radius for the simulated alpha-viscoties.
The cases of $\alpha=0.01$, $\alpha=0.005$, and $\alpha=0.001$ do not show any significant spin-up of the star after 3/4 of a Keplerian rotation, whence there is not significant shear connection between the disk and the stellar surface. Therefore disk matter retains its Keplerian motion much closer to the star which means that even though the accretion rates are smaller, significant amounts of the dissipation can still be generated in the BL close to the star.

\subsection{Formation of gravity waves}\label{subsec:grav}
For the $\alpha=0.1$ case, the edge of the boundary layer ($d\Omega/dr\equiv 0$, see \cite{Frank02}) is located at $1.06R_*$. At $R=1.03R_*$, $v_\phi$ is about 92\% of $v_K$. This means that only 84\% of the gravitational pull is supported by the centrifugal force while the rest is supported by pressure. 
\begin{figure}[thp]
 \epsfig{file=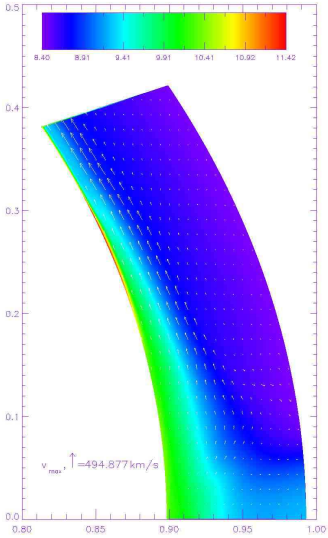, width=0.6\linewidth}
 \caption{A contour plot of the pressure for the $\alpha=0.1$ case after 3/4 of a Keplerian rotation period shown with a logarithmic scale.}\label{fig:pressure91}
\end{figure}
The pressure for $\alpha=0.1$ is illustrated in fig.~\ref{fig:pressure91} and comes from the build up of density due to the high accretion rate facilitated by the high viscosity which drags down material from the disk and also keeps it from moving rapidly towards the poles once it makes contact with the WD surface. 
This results in a dense band at the foot point of the disk which causes a gravity wave of surface matter to spread towards the poles. The matter inflow is, however, sufficiently large to cause the disk material to overflow the gravity wave supersonically as described above and shown in fig.~\ref{fig:omega91}. 
Over time the thick band will spin up the surface and thus prevent shear forces from accreting matter since the differential rotation decreases as the BL is spun up.

The propensity for the creation of gravity waves and their magnitude decreases rapidly with lower values of $\alpha$. Even for $\alpha=0.03$, the effect is only marginal and for lower values it is no longer present. The effect is thus only present during dwarf novae in the outburst stage. Furthermore the gravity wave might be transient as the surface adjusts to a fluctuations in the accretion rate. For lower values of $\alpha$, the matter accretes slowly inwards and spreads towards the poles in a uniformly thick layer.

\subsection{Dissipative heating and optical depth of the BL}\label{subsec:optical}
Observationally, the high density of the spreading BL leads to optically thick conditions for $\alpha=0.1$ and $\alpha=0.03$ in a band extending at least 30 degrees away from the disk's centerline. The band should therefore provide a continuum component to the observed spectrum. Here we use the criterion that $-\int_\infty^0 \kappa \rho_{disk}dr>2/3$, where $\kappa=0.34\,\textrm{g}\,\textrm{cm}^{-3}$ is the Thomson scattering opacity of a fully ionized solar composition. This constitutes the minimum amount of scattering and thus provides a lower bound of the opaqueness of the matter. 
Since the density is lower for lower values of $\alpha$, the BL is optically thin more than 15 degrees away from the disk itself for $\alpha=0.01$, $\alpha=0.005$, and $\alpha=0.001$.

\begin{figure}[thp]
 \epsfig{file=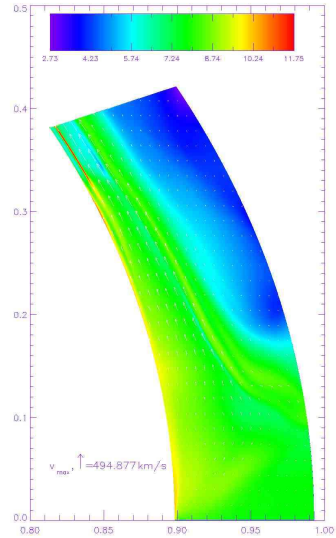,width=0.6\linewidth}
 \caption{A logarithmic contour plot of the dissipation rate for the $\alpha=0.1$ case after 3/4 of a Keplerian rotation period.}\label{fig:diss91}
\end{figure}

\begin{figure}[thp]
 \epsfig{file=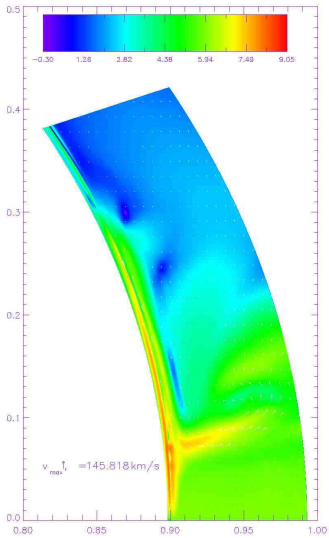,width=0.6\linewidth}
 \caption{A logarithmic contour plot of the dissipation rate for the $\alpha=0.001$ case after 3/4 of a Keplerian rotation period. Notice that the color scale is different for fig.~\ref{fig:diss91}.}\label{fig:diss95}
\end{figure}

The local rate of energy dissipation depends on the divergence of the angular velocity \cite{Mihalas84}, so fig.~\ref{fig:vlza} also indirectly shows where heat is released and thus where radiation is emitted. 
Fig.~\ref{fig:diss91} shows that the supersonic impact of the accretion flow following the cusp dissipates the most energy. However, energy is also dissipated at the interface between the BL and the surface. This is better seen in fig.~\ref{fig:diss95} which shows that the toroidially rotating matter dissipates rotational kinetic energy along its entire interface with the surface as it slides against it. This may explain the optically thin BL which may explain the line emission as seen in observations of U Gem during the quiescent phase \cite{Fisker05c}.

\section{Conclusion}
We have numerically simulated the structural dynamics of the BL for an accreting white dwarf surrounded by an $\alpha$-disk for different values of $\alpha$.
The structure and dynamics of the BL is important, because it determines the specifics of the radiated spectrum emitted from this region which may account for up to half of the total energy released in the accretion process. 
For high values of $\alpha$, the BL is optically thick and extends more than 30 degrees to either side of the disk plane after 3/4 of a Keplerian rotation period ($t_K$=19s).  

The dynamics of the accretion process in the boundary layer is also important, because it determines the mixing ratio between the accreted matter and the underlying surface matter and thus determines the initial composition which is important for nucleosynthesis in classical novae. 
Here the simulations show that high values of $\alpha$ result in a spreading BL which sets off gravity waves in the surface matter, where the accretion flow moves supersonically over the cusp making it susceptible to the rapid development of gravity wave and/or Kelvin-Helmholtz instabilities.  
The low viscosity cases also show a spreading BL, but here the accretion flow does not set off gravity waves and it is optically thin.

The cooling time for the plasma is very short for all the simulations \cite{Sutherland93}, so in an upcoming paper by Balsara \& Fisker we repeat the calculations without a viscous dissipation in the energy equation. This corresponds to an instant steady state cooling of the layer. Incorporating a flux limited diffusion radiation transport will be defered to a later paper. 

\begin{ack}                               
JLF was supported by NSF-PFC grant PHY02-16783 through the Joint Institute of Nuclear Astrophysics. DSB was supported by NSF AST-0132246 and DMS-0204640. TB thanks the REU program at Notre Dame for support via NSF PHY-0097624. This work used the HPCC cluster at the University of Notre Dame and was partially supported by the National Center for Supercomputing Applications under AST050031 and utilized the NCSA SGI Altix.
\end{ack}

\bibliographystyle{plain}        



\end{document}